\documentstyle [aps,prl,epsf,twocolumn]{revtex}
\begin{document}
\twocolumn[\hsize\textwidth\columnwidth\hsize\csname@twocolumnfalse%
\endcsname

\title{
Wavefunction and level statistics of random two dimensional gauge fields
}

\author{ J. A. Verg\'es}
\address{
Instituto de Ciencia de Materiales de Madrid,
Consejo Superior de Investigaciones Cient\'{\i}ficas,\\
Cantoblanco, E-28049 Madrid, Spain}

\date{Received 27 June 1995}

\maketitle

\begin{abstract}
Level and wavefunction statistics have been studied for two dimensional
clusters of the square lattice in the presence of random magnetic fluxes.
Fluxes traversing lattice plaquettes are distributed
uniformly between $- {1 \over 2} \Phi_0$ and ${1 \over 2} \Phi_0$
with $\Phi_0$ the flux quantum.
All considered statistics start close to the corresponding Wigner-Dyson
distribution
for small system sizes and monotonically move towards Poisson statistics
as the cluster size increases.
Scaling is quite rapid for states close to the band edges but really difficult
to observe for states well within the band.
Localization properties are discussed considering two different scenarios.
Experimental measurement of one of the considered statistics
--wavefunction statistics seems the most promising one--
could discern between both possibilities.
A real version of the previous model, i.e., a system that is invariant
under time reversal, has been studied concurrently to
get coincidences and differences with the Hermitian model.
\end{abstract}

]

\section{Introduction}

There is a general belief based on Anderson localization
theory\cite{localizacion}
that all states of two-dimensional (2D) systems are localized in the
absence of a magnetic field. The situation is not so clear when time-reversal
symmetry is destroyed by a magnetic field.
The motion of a single particle in a 2D random magnetic field is
attracting strong interest since it is related both to the half-filled
Quantum Hall Effect\cite{kalmeyer,halperin}
and the slave-boson description of high $T_c$
superconductors\cite{slave}.
The existence of recent experiments measuring 
transport properties in a static random magnetic
field adds considerable interest to this subject\cite{exp}.

While perturbative renormalization group calculations show that all
states are localized\cite{hikami,mirlin}, the eventual presence of an extra
term could give rise to a Kosterlitz-Thouless transition from the localized
phase to a phase with power law correlations\cite{zhang}.
Numerical simulations have used a variety of forms:
the iterative strip method has been used by Sugiyama and Nagaosa\cite{nagaosa},
Avishai {\it et al.}\cite{kohmoto}, Liu {\it et al.}\cite{sarma},
and Kalmeyer {\it et al.}\cite{kalmeyer},
analysis of participation ratio has been employed by Pryor and Zee\cite{pryor}
and Kalmeyer and Zhang\cite{kalmeyer},
tails in the density of states have been discussed by Gavazzi {\it et
al.}\cite{wheatley} and Barelli {\it et al.}\cite{barelli},
a network model has been introduced by Lee and Chalker\cite{dkklee},
the diffusion of electrons has been studied by solving the time-dependent
Schr\"odinger equation\cite{kawa},
and finally, Hall conductivity has been used as a way for determining
energy regions where extended states dominated in the thermodynamic
limit\cite{sheng}.
Loosely speaking, states {\it look} extended in
finite samples as soon as the edge of the band is left
and a definite answer is quite difficult.
Therefore, the most extended conclusion is  that a  mobility edge
separates localized states from extended or critical states.
On the contrary,
Sugiyama and Nagaosa\cite{nagaosa} conclude that {\it all} states are localized
in random magnetic fields based on the non-existence of a second branch
in the scaling function whereas the same conclusion is reached by
Lee and Chalker\cite{dkklee} based on finite values of the
localization length obtained in the semiclassical limit described
by their model.

In this work, I present numerical evidence showing that,
in the presence of random fluxes, states scale towards random (Poisson)
statistics everywhere in the band. Several standard statistics
(wavefunctions amplitude statistics, nearest neighbor spacing statistics
and number variance) are used in conjunction with other criteria
(for example, spatial
extension of wavefunctions as given by the participation ratio)
to show a systematic tendency towards Poisson statistics (the one that
characterizes systems with a spectrum formed by exponentially localized states)
as the size of the random samples increases.
Scaling is quite rapid near the band edge where states
"look" exponentially localized whereas it seems to follow a logarithmic
law near the band
center where states "look" fractal\cite{fractal}.
Nevertheless, since my analysis is based on statistical magnitudes, the
eventual existence of a countable set of extended states can not be disproved.
In other words, the results presented in the next Sections prove that
states scale towards localization {\it on average}.

Recent results regarding level statistics can be found in the
literature\cite{barelli,stat}. Most of them deal with
the characterization of a new universal distribution function
that would be appropriate at the metal-insulator transition.
As the results presented in the rest of the paper show,
signs suggesting the existence of a critical energy (mobility edge)
at which scaling changes sign have not been found.
{\it Scaling flows in the same direction in all energy regions}.
The paper is organized as follows. Section II defines the model that has been
numerically solved in the paper,
Section III gives the main results grouped under
three subjects (Density of states, wavefunctions and level statistics) and
finally, main conclusions are collected in the last Section.

\section{Lattice model}

The Hamiltonian describing random gauge fields on a $L \times L$ cluster
of the square lattice is :

\begin{equation}
\hat H = - ~t {\sum_{<l l'>}}
        {e^{2 \pi i \phi_{l l'}}
        \hat c_{l}^{\dag} \hat c_{l'}}
        ~~,
\label{RMF}
\end{equation}

\noindent
where $\hat c_{l}^{\dag}$ creates an electron on site $l$,
$l$ and $l'$ are nearest-neighbor sites, and
$- t$ is the hopping energy (hereafter $t=1$ is chosen).
The flux through a given loop $S$ on the lattice is
$\Phi_S = \sum_{<l,l'> \in S} \phi_{l l'}$
measured in units of flux quantum $\Phi_0 = {{hc} \over e}$.
Link fields satisfy $\phi_{l l'}=-\phi_{l' l}$.
Although some checks have been performed for the Meissner phase\cite{wheatley}
(link fields uniformly, randomly distributed in the interval
$[- {1 \over 2},{1 \over 2}]$),
the bulk of the calculations have been done for the Debye phase in which
uncorrelated fluxes are randomly selected from the interval
$[- {1 \over 2},{1 \over 2}]$.
Notice that in any case,
the situation of maximum possible disorder is considered except when
the density of states is analysed as a function of disorder.
Model (\ref{RMF}) will be called
RMF (Random Magnetic Fluxes or Random Magnetic Field) model
in the rest of the paper.

In order to recover standard results, I have always run checks
for a {\it real} version of Hamiltonian (\ref{RMF}).
To this end, link
field values are restricted to $0$ and $1/2$ and chosen randomly. The
model describes a square lattice in which the hopping integral is constant
in absolute value but takes random signs.
This model will be shortly mentioned as RHS (Random Hopping Signs) model.
Standard localization theory should apply to this model.
Both models are equivalent for a particular choice of random fluxes;
when $\phi_{l l'}$ is always equal to $1 \over 4$ but has random sign,
hopping integrals take random $\pm i$ complex values.
Thanks to the bipartite character of the square lattice, this Hamiltonian
can be made real: it suffices a canonical transformation
of all operators on one of the sublattices from $\hat c_l$ to $i \hat c_l$.
In this way, this particular complex Hamiltonian is described by
the same matrices as the RHS model.
This case provides an overlap between real random
Hamiltonians that according to standard theory show complete localization
and a Hamiltonian with complex eigenfunctions.

\section{Results}

\subsection{Density of states}

One of the easiest ways allowable to characterize the effect of disorder
is the density of states (DOS):

\begin{equation}
N(E)={1 \over N} \sum_{\alpha} \delta(E-E_{\alpha})
\end{equation}

\noindent
where the sum runs over all eigenstates of the system.
For arbitrary disorder $\{ \phi_{l l'} \}$ the eigenvalues of
Hamiltonian (\ref{RMF}) lie in the interval $[-4,4]$. Moreover, there is
also an exact symmetry on a bipartite lattice like the square lattice.
For every eigenfunction of energy $E$ there is an eigenfunction of
energy $-E$ whose amplitudes have opposite sign on one of the sublattices.
As a consequence, the density of states is symmetric about $E=0$.
The same symmetry forces the existence of an eigenstate of zero energy
in clusters of an odd number of sites.

\begin{figure}
\begin{picture}(236,400) (-50,0)
\epsfbox{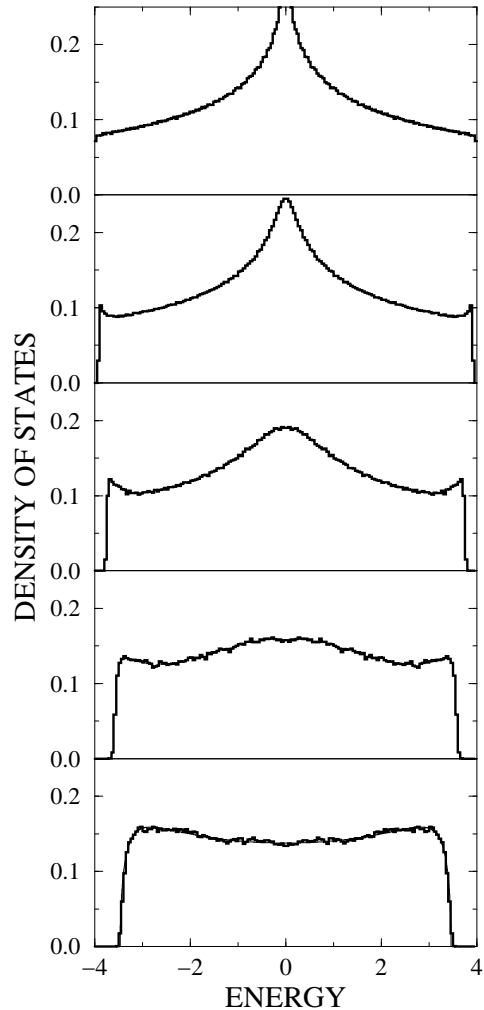}
\end{picture}
\caption{Evolution of the density of states as the interval
$[-\phi_{\rm max},\phi_{\rm max}]$ from which random values of the flux
are selected increases.
From top to bottom: ideal system, $\phi_{\rm max}={1 \over 8}, {1 \over 4},
{3 \over 8}$ and ${1 \over 2}$. The DOS of a Bethe lattice of coordination
equal to four has been plotted (continuous line)
in the bottom panel to allow comparison with the RMF result.}
\label{dos}
\end{figure}

Generally, an average
over disorder realizations is done in order to get a smooth result.
Averaging is particularly important when Anderson model is studied
for large disorder because band width increases and the number of states per
energy unit diminish.
Since the disorder given by Hamiltonian (\ref{RMF}) systematically decreases
the width of the band, the study of only one large sample suffices to get
nice results. Fig.(\ref{dos}) shows the evolution of the density of
states obtained for a $200 \times 200$ system as disorder (random fluxes
distributed in the interval $[-\phi_{\rm max},\phi_{\rm max}]$) increases.
A small amount of disorder destroys the logarithmic singularity at the
center of the regular system and shrinks somewhat the band.
Peaks appear near
the band edges due to some kind of accumulation of states due to the band
skrinkage. Finally, a smooth curve is obtained for the maximum possible
amount of disorder: random fluxes between $-{1 \over 2}$ and $1 \over 2$.
In this case, the density of states of a Bethe lattice of coordination equal
to four:

\begin{equation}
N(E)={{2 \over \pi} {{\sqrt{12-E^2}} \over {16-E^2}}}
\end{equation}

\noindent
gives a very good approximation to the random result (it is difficult to
distinguish between histogram and continuous curve in the bottom panel
of Fig.(\ref{dos})).
Although the characterization and deep understanding of the peak appearing
close to the edges at small disorder deserves further study, the rest of the
paper is devoted exclusively to the case of maximum disorder (bottom panel of
Fig.(\ref{dos})).

\begin{figure}
\begin{picture}(236,275) (20,-20)
\epsfbox{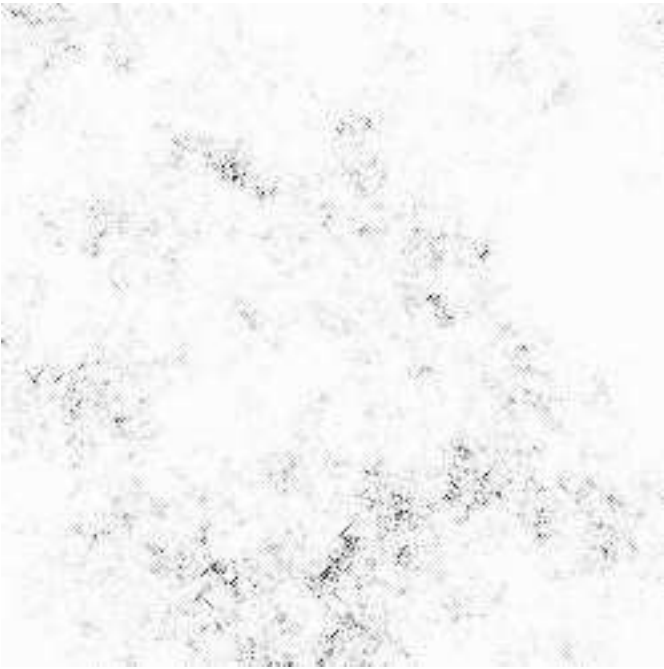}
\end{picture}
\caption{Typical "extended" wavefunction of a $250 \times 250$ cluster and
energy $E \approx -\pi/50$.
\label{47}}
\end{figure}

\subsection{Wavefunctions}

Wavefunctions can be obtained using standard algorithms for Hermitian 
band matrices\cite{schwarz}
if the size of the matrix is not too large (a typical workstation can
deal with a $200 \times 200$ cluster, i.e., with a $40000 \times 40000$ matrix
if only eigenvalues are calculated but only with a $64 \times 64$
system (a matrix of dimension equal to 4096) if eigenvectors should be
obtained). On the other hand,
inverse iteration\cite{recipes} allows the calculation of
particular eigenvectors. The algorithm starts with some
random values for the components of the wavefunction $\psi_0$
of energy $E$ and proceeds as follows:

\begin{equation}
\begin{array}{c}
{\psi_1} = (E - \hat H)^{-1} {\psi_0} \\
{\psi_2} = (E - \hat H)^{-1} {\psi_1} \\
... \\
\end{array}
~~~~~~~~~~~.
\end{equation}

\noindent
Convergence is reached when $|\psi_i-\psi_{i-1}|$ is smaller than some
tolerance $\epsilon$ ($\epsilon = 10^{-8}$ works fine in double precision).
Energy should be updated after some number of cycles (about 10):
$$
E_{i}=<\psi_{i}|\hat H|{\psi_{i}}> ~~~.
$$

\begin{figure}
\begin{picture}(236,275) (20,-20)
\epsfbox{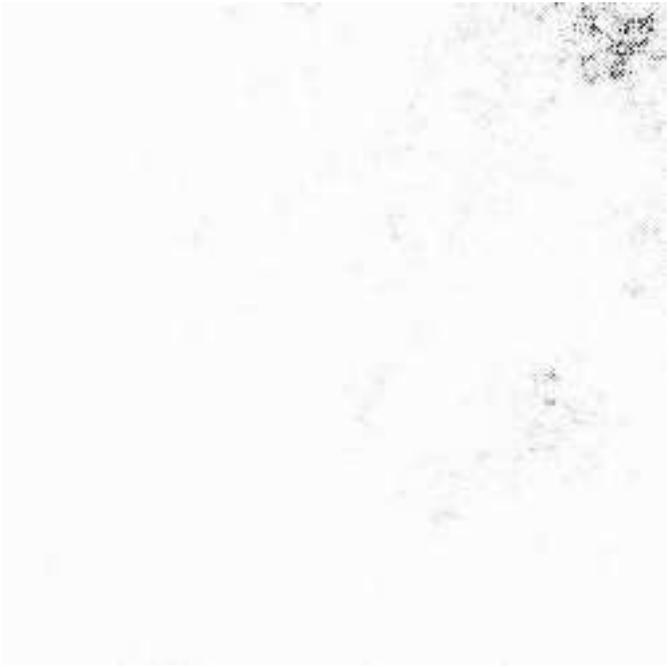}
\end{picture}
\caption{Typical "localized" wavefunction of a $250 \times 250$ cluster and
energy $E \approx -\pi/50$.
\label{49}}
\end{figure}

Taking advantage of the band structure of the Hamiltonian, eigenstates
for clusters as large as $300 \times 300$ can be obtained using standard
linear algebra subroutines. Sometimes larger systems are
accesible: for example, since $E=0$ is always an eigenenergy of Hamiltonian
(\ref{RMF}) when the number of sites is odd, total number of iterations is
greatly reduced and eigenvectors for $399 \times 399$ clusters can be
obtained. The computational thresholds that have been given
refer to complex Hermitian matrices. Somewhat larger systems can
be studied for the real version of model (\ref{RMF}).

Wavefunctions produced by Hamiltonian (\ref{RMF}) present a great richness.
We get {\it bona fide} exponentially localized wavefunctions for energies close
to the band edges, almost localized wavefunctions
(i.e., wavefunctions that show non
vanishing amplitudes in a small part of the cluster) in any part of the
energy spectrum, quasi one-dimensional eigenstates in many cases and also,
quasi extended wavefuntions in the main part of the spectrum.
Figs.(\ref{47}) and(\ref{49}) show typical results at cluster sizes at which
only inverse iteration allows the calculation of specific
eigenstates\cite{recipes}.
Wavefunctions are represented through a 256 values grey scale
in which darker means larger values of ${|\psi_l|}^2$ being $\psi_l$ the
wavefunction amplitude on site $l$.
Although a large collection of such wavefunctions have been compiled,
it is clear that the simple display of them does not help to the analysis of
localization. Therefore, I have use standard tools to analyse
their extension and the way they behave as the size of the sytem is increased.

\subsubsection{Participation ratio}

A good measure of the spatial extension of an eigenstate is given by the
participation ratio:

\begin{equation}
P = \biggl( {\sum_{l=1}^{L \times L}} {{|\psi_l|}^4} \biggr)^{-1}
        ~~,
\label{P}
\end{equation}

\noindent
where $\psi_l$ is the amplitude of wavefunction on site $l$.
$P$ can be understood as the number of sites covered by the wavefunction and
reaches a maximum finite value for localized states.
In order to avoid systematic errors due to the use of a biased definition,
I have also tried alternative definitions of the spatial extension covered by a
wavefunction. One is given in Casati and Molinari work\cite{casati}:

\begin{equation}
C=\exp \left( -{\sum_{l=1}^{L \times L}} {{|\psi_l|}^2} \log({|\psi_l|}^2)
         \right) ~~~,
\label{C}
\end{equation}

\noindent
a definition of the number of sites asociated to a wavefunction that
emphasizes small amplitudes and gives consequently larger values than
the previous definition. Still another definition is based in the basic
understanding of ${|\psi_l|}^2$ as the probability of finding the particle 
at site $l$. We proceed in two steps: (i)
weights ${|\psi_l|}^2$ are ordered in decreasing
magnitude and (ii), $S$ is defined as the number of ordered weights
that have to be added to reach some percentage of the total weight 1 of
the state. I have tried different percentages to learn that this simplest
definition gives results for $S$ that are practically identical to $C$ when
$99 \%$ of the total weight is considered. Using $90 \%$ as the cutoff,
results closely follow those obtained using the standard $P$ definition.
This knowledge gave me further confidence in the reliability of the usual
definitions of number of sites covered by a wavefunction.

\begin{figure}
\begin{picture}(236,350) (-70,-50)
\epsfbox{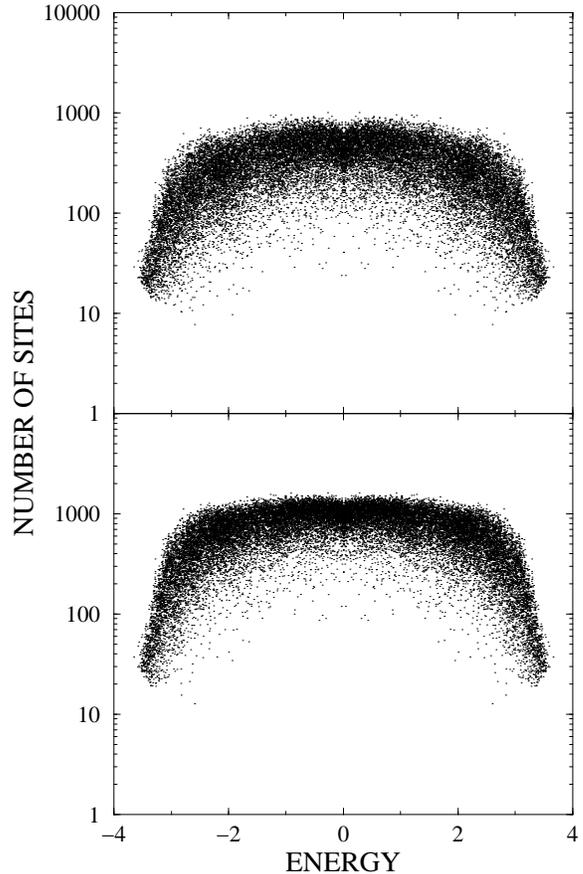}
\end{picture}
\caption{Number of sites for all states of seven samples of $64 \times 64$ size.
Top: {\em standard} definition of the participation ratio(\ref{P}). 
Bottom: Logarithmic definition(\ref{C}).
\label{sites_whole_rhs}}
\end{figure}

\begin{figure}
\begin{picture}(236,350) (-70,-50)
\epsfbox{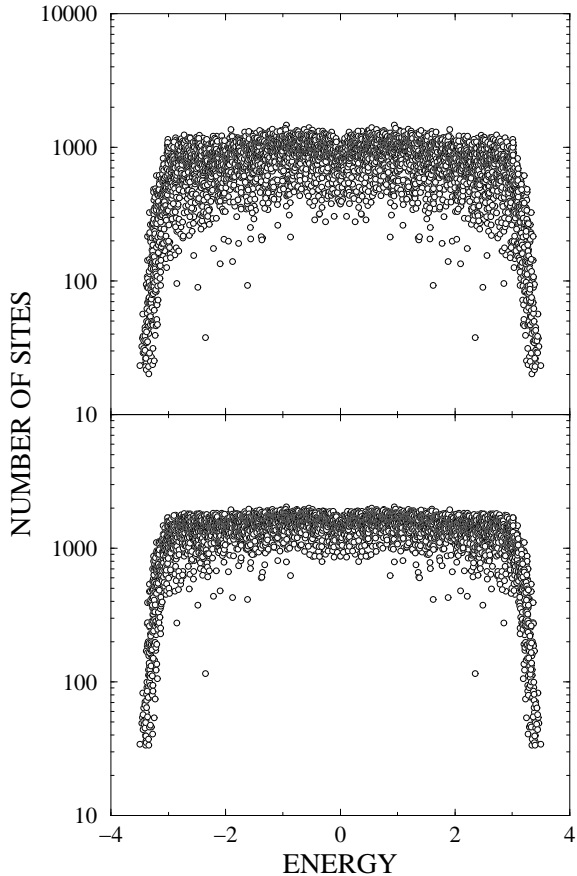}
\end{picture}
\caption{Number of sites for all states of two samples of $58 \times 58$ size.
Top: {\em standard} definition of the participation ratio(\ref{P}). 
Bottom: Logarithmic definition(\ref{C}).
\label{sites_whole_rmf}}
\end{figure}

Fig.(\ref{sites_whole_rhs}) and Fig.(\ref{sites_whole_rmf}) show both
$P$ and $C$ for a collection of the largest clusters for which
the complete set of eigenergies and eigenvectors can be obtained using my
limited computing facilities.
Schwarz algorithm for symmetric or Hermitian
band matrices has been used\cite{schwarz}. These results show that spatial
extension is small near band edges but increases quickly towards the band
center. In fact, the behavior is really sharp for the complex RMF model.
One can be tempted to speak about some kind of pseudo-mobility edge at
these energies ($\sim \pm 3$) where spatial extent sharply increases.
In fact these energies
coincide with the values given in the literature as the limits of a reliable
calculation of the localization length (see, for example, the paper by
Sugiyama and Nagaosa\cite{nagaosa}). Moreover, while states well within
the band show a steady power law increase in their spatial extension,
states close
to the edges {\em do not} extend as the size of the system is increased.

\begin{figure}
\begin{picture}(236,295) (14,20)
\epsfbox{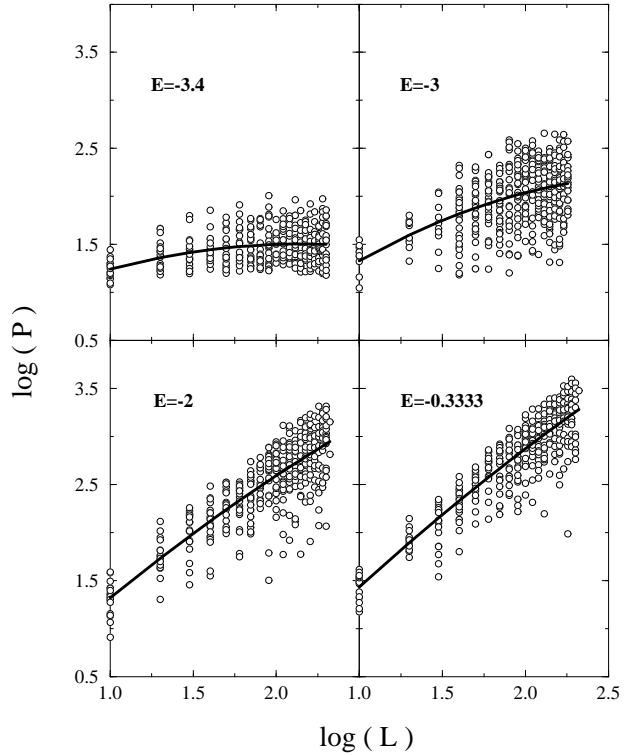}
\end{picture}
\caption{Spatial extension of eigenstates produced by the real version of
Hamiltonian (\ref{RMF}) at different energies.
Circles give the value of $ \log (P) $ of
the state that is closest to the selected energy
for each sample of a set of realizations of disorder.
$L$ is the cluster side.
Solid lines are quadratic fits to the log-log values.
\label{sites_rhs}}
\end{figure}

\begin{figure}
\begin{picture}(236,290) (0,10)
\epsfbox{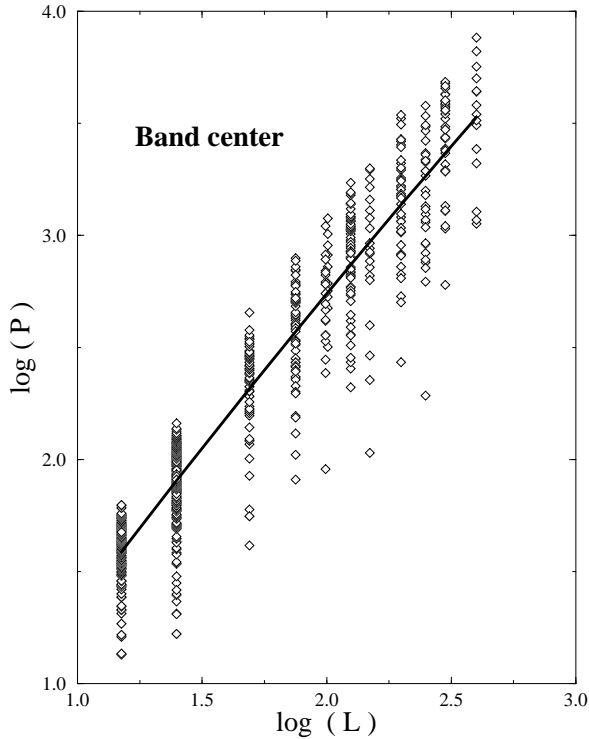}
\end{picture}
\caption{Spatial extension of eigenstates produced by
Hamiltonian Eq.(\ref{RMF}) at the band center. Diamonds correspond
to different realizations of disorder for several sizes ($L \times L$ cluster).
Solid line is a quadratic fit to the log-log values.
\label{sites_rmf}}
\end{figure}

The scaling behavior of the spatial extension of states 
has been carefully analysed at particular energies.
In this analysis inverse iteration is used and therefore, eigenstates
for much larger systems are available.
Fig.(\ref{sites_rhs}) collects results for RHS model at several energies
while Fig.(\ref{sites_rmf}) shows spatial extension results for RMF model just
at the band center. Let me begin the discussion with some comments about the
results displayed in Fig.(\ref{sites_rhs}).
Several energies covering different parts of the spectrum have been chosen.
While it is clear that the growth of the spatial extension
is bound in the lower part of the band ($E=-3.4$ and perhaps $E=-3$),
results at energies well within the band show eigenstates that
extend more and more as the system size increases. Of course, the
validity of this conclusion is limited by
the sizes that are reachable by a numerical study
(the largest cluster analysed in this case is $210 \times 210$).
Nevertheless, a quadratic
fit to the log-log plot shows a small negative curvature that
indicates that after a region in which $<P> \sim L^\alpha$
($\alpha \leq 2$), a region of true exponential localization of wavefunctions
could follow. Again a word of caution is in order at this point:
the inference of the asymptotic behavior of
coverage from the analysis of a limited range of sizes is really a hard
question in any numerical study. Therefore, the only conclusion that seems
to be sure is that coverage increases following a power law with
an exponent smaller than 2 {\it at the most}.

A similar analysis is displayed in Fig.(\ref{sites_rmf}) for the Hermitian
Hamiltonian. Participation ratios for states at the band center are given
for increasing cluster sizes. In that case, the maximum size allowed by our
computing capabilities is $399 \times 399$.
Even for this system size, the number
of sites covered by eigenstates at the band center is following on average
a law close to $<P> \sim L^{1.6}$. The curvature of a
quadratic fit to the log-log data shows again negative --although really
small-- curvature (solid line of Fig.(\ref{sites_rmf})).
So, we are forced to conclude that numerical methods based in the study
of the scaling properties of individual states are not enough
to show exponential localization in 2D random gauge fields.

\subsubsection{Fractal dimension}

Fractal dimension has been defined as:

\begin{equation}
\alpha ={{d \log (P)} \over {d \log (L)}} ~~~~~ .
\end{equation}

\noindent
It characterizes energies or system sizes for which the number of sites
covered by eigenstates
increases following a power law (Number of sites $\propto L^\alpha$).
It has been calculated for the whole spectrum using energy averaged
values for the number of sites and two different system sizes to calculate
the derivative. In other words, we have:

\begin{equation}
\alpha ={{\Delta \log <P>} \over {\Delta \log <L>}} ~.
\end{equation}

\begin{figure}
\begin{picture}(236,210) (-30,-70)
\epsfbox{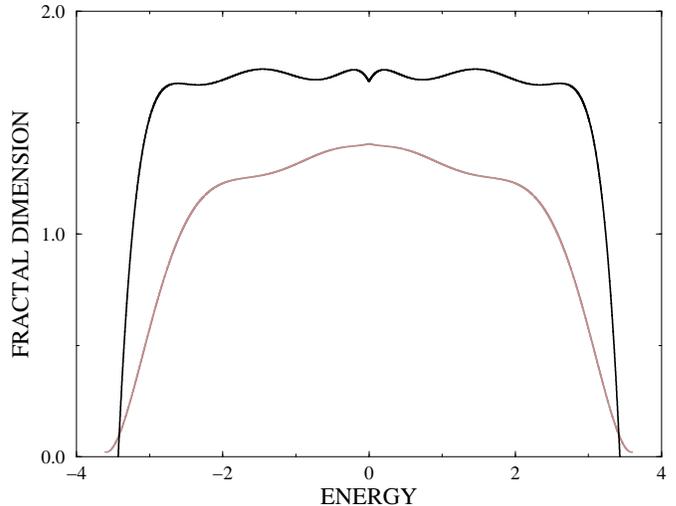}
\end{picture}
\caption{Fractal dimension of both models (RHS: grey line; RMF: black line)
as a function of the position of states within the band.
\label{dimension}}
\end{figure}

Fig.(\ref{dimension}) gives our numerical results for both models.
Sizes $L=32, 64$ have been used for RHS model whereas sizes $L=24, 48$
did the work for the more difficult RMF model. It can be seen that although
the overall behavior is similar for both models, the increase to
$\approx 1.8$ is so sharp for the RMF model that the idea of some critical
energy separating exponentially localized states from states following
a power law decay could be considered.

The fact that fractal dimension for averages of the coverage $P$
is below $1.8$,
together with the more detailed results at particular energies shown in
the previous subsubsection, proves overall localization {\it at least
in a power law way}:

$$
{<P> \over {L^2}} \rightarrow 0 ~~~~~.
$$

\noindent
This behavior is in striking contrast with the result that
applies to wavefunctions of the GUE:

$$
{<P> \over N} \rightarrow {1 \over 3} ~~~~~,
$$

\noindent
where the total number of sites $N$ plays the role of $L^2$ in the previous
expression. For the same reason, localization is also present in
the simplified RHS model for which fractal
dimension is always less than $1.4$. Actually, states are less extended for
this model.

\subsubsection{Wavefunction amplitude statistics}

Wavefunction statistics measures the spatial variation of eigenfunctions
over the system under study.
Crystalline wavefunction have constant amplitude $1/N$ whereas wavefunction
of disordered or chaotic systems show large fluctuations.
The wavefunction statistics of the canonical Gaussian ensembles is given by the
corresponding Porter-Thomas distribution\cite{poto}. In particular, we have

\begin{equation}
f(t)={1 \over {\sqrt{2 \pi t}}} \exp(-t/2) ~~~~~~~,
\end{equation}

\noindent
for the Gaussian Orthogonal Ensemble, where $t$ is the squared wavefunction
amplitude divided by its mean value, i.e.,
($t=N {\psi (\bf r)}^2$ ) being $N$ the number of sites.
Graphical representation is improved if variable $x=\log (t)$ is used instead
of $t$. We get:

\begin{equation}
g(x)={1 \over {\sqrt{2 \pi}}} \exp \left\{ {1 \over 2} [x-\exp(x)] \right\} ~~~.
\label{ptgoe}
\end{equation}

\noindent
On the other hand, we have:

\begin{equation}
f(t)=\exp(-t) ~~~~~~~,
\end{equation}

\noindent
for the Gaussian Unitary Ensemble, where $t$ is the squared wavefunction
amplitude divided by its mean value, i.e.,
($t=N {\psi (\bf r)}^* \psi (\bf r)$).
Doing the same variable transformation as before, we get:

\begin{equation}
g(x)=\exp(x-\exp(x)) ~~~~~~~.
\label{ptgue}
\end{equation}

\begin{figure}
\begin{picture}(236,200) (-30,-70)
\epsfbox{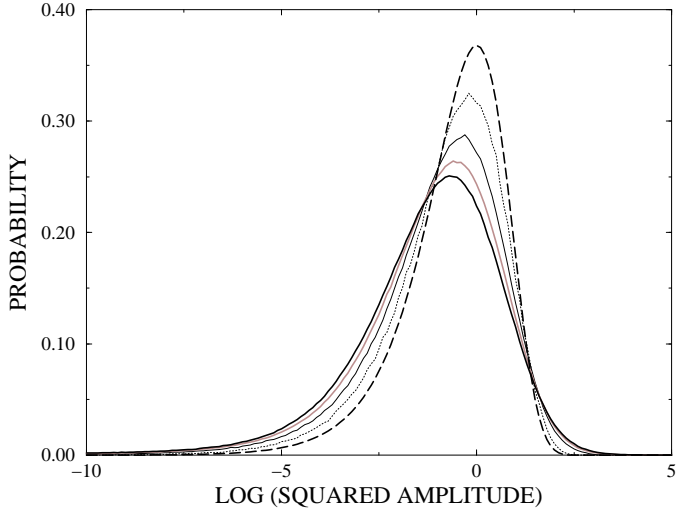}
\end{picture}
\caption{ Wavefunction statistics of RMF model.
Distribution of wavefunction weight for
i) an average over all states of 256 $8 \times 8$ samples (dotted line),
ii) an average over all states of 64 $16 \times 16$ samples (solid line),
iii) an average over all states of sixteen $32 \times 32$ samples
(grey thick solid line) and,
iv) an average over all states of two $58 \times 58$ samples
(black thick solid line).
GUE Porter-Thomas distribution is given for comparison (dashed line).}
\label{pt_rmf_did}
\end{figure}

Deviations from these canonical distributions have been recently
calculated by Falko and Efetov\cite{efetov}. Their work applies to
pre-localized states in disordered conductors. Numerically, it is
a simple task getting deviations from universal behavior.
Fig.(\ref{pt_rmf_did}) shows the rapid deviation of RMF wavefunction
statistics from the corresponding Porter-Thomas distribution (Eq.(\ref{ptgue}))
as soon as eigenfunctions of medium-size matrices are considered.
In this calculation, all states --regardingless their energy-- have been used
to compute the probability density. Close to the maximum capability
of my computer ($58 \times 58$ clusters), it can be seen that the
ratio of evolution of wavefunction statistics with size is decreasing.

\begin{figure}
\begin{picture}(236,200) (-30,-70)
\epsfbox{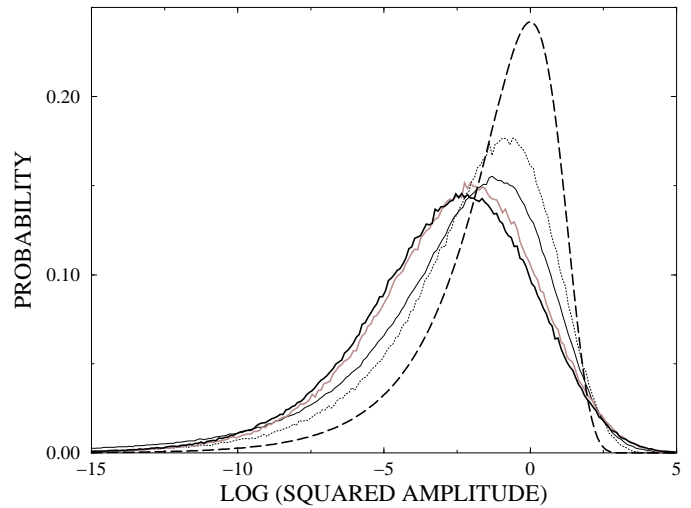}
\end{picture}
\caption{ Wavefunction statistics of RHS model.
Distribution of wavefunction weight for
i) an average over all states of sixteen $16 \times 16$ samples (dotted line),
ii) an average over all states of four $32 \times 32$ samples (solid line),
iii) an average over fifteen states of energy $E \approx -\pi/50$ obtained
for different $200 \times 200$ clusters (grey thick solid line) and,
iv) an average over ten states of energy $E \approx -\pi/50$ obtained
for different $300 \times 300$ clusters  (black thick solid line).
GOE Porter-Thomas distribution is given for comparison (dashed line).}
\label{pt_rhs}
\end{figure}

\begin{figure}
\begin{picture}(236,200) (-30,-70)
\epsfbox{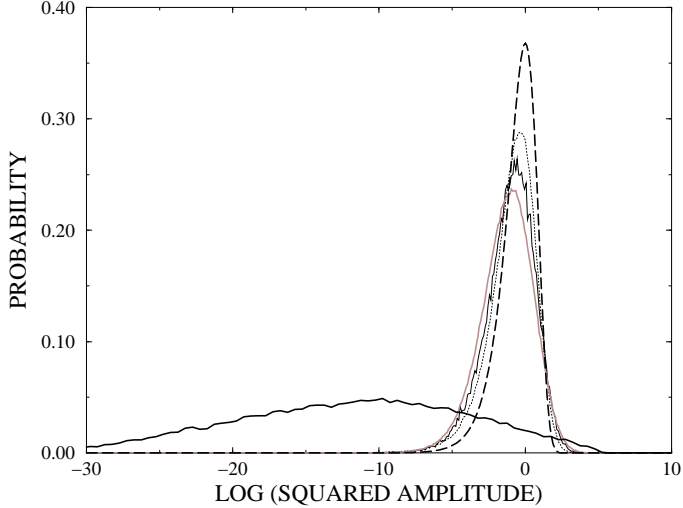}
\end{picture}
\caption{ Wavefunction statistics of RMF model.
Distribution of wavefunction weight for
i) an average over all states of sixteen $16 \times 16$ samples (dotted line),
ii) an average over forty states of energy $E \approx -\pi/50$ obtained
for different $50 \times 50$ clusters (solid line),
iii) an average over six states of energy $E \approx -\pi/50$ obtained
for different $300 \times 300$ clusters (grey thick solid line) and,
iv) an average over twenty states of energy $E \approx -3.4$ obtained
for different $50 \times 50$ clusters  (black thick solid line).
GUE Porter-Thomas distribution is given for comparison (dashed line).}
\label{pt_rmf}
\end{figure}

The divergence of wavefunction statistics from the corresponding Porter-Thomas
distributions shown in the last Figure is confirmed by the calculation
of wavefunctions of much larger clusters close to a fixed energy (I have used
inverse iteration starting at $E=-\pi/50$ trying to get unbiased eigenstates
that could exist at special energies).
Fig.(\ref{pt_rhs}) shows selected results for RHS model while
Fig.(\ref{pt_rmf}) gives RMF results. In both cases, the same clear tendency
of diverging from Porter-Thomas statistics as size increases, can be seen.
Nevertheless, the
distance to a {\it typical} statistics produced by exponentially localized 
wavefunctions is still large (thick solid line of Fig.(\ref{pt_rmf}) shows
wavefunction statistics at the band edge). Notice that exponential
localization implies a statistics that shifts towards left (smaller weights
are more probable) as the size of the system increases
beyond the localization length.

\subsection{Level statistics}

\subsubsection{Nearest neighbor spacing statistics}

Level statistics has been recently used to characterize the properties
of spectra near the mobility edge\cite{stat}. It is
well known that the distribution of nearest neighbor spacings is of
the Wigner-Dyson type\cite{porter}
when particles move through the whole slightly disordered sample
and changes to Poisson when states become exponentially localized.
Simplifying the argument,
we can say that only states localized in different spatial regions
do not interact through the Hamiltonian and are, therefore, allowed to lie
at the same energy.
According to this theory, level statistics should move towards Poisson
as the size of the system increases {\it whenever} states are localized
in the thermodynamic limit whereas it should move towards the corresponding
Wigner-Dyson distribution if states are extended in the infinite
system\cite{ls}.

\begin{figure}
\begin{picture}(236,270) (0,-30)
\epsfbox{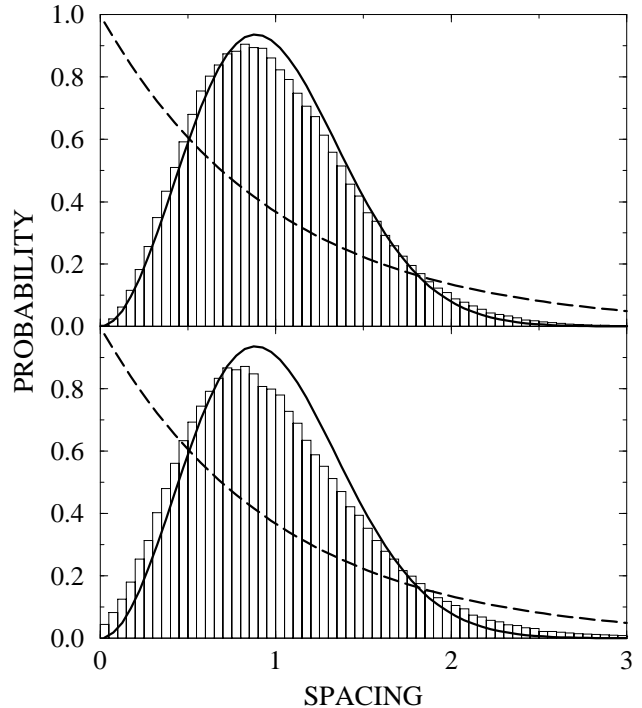}
\end{picture}
\caption{Top panel:
Distribution of nearest neighbor spacings obtained for 4096 randomly
generated $12 \times 12$ clusters (histogram).
Bottom panel:
Distribution of nearest neighbor spacings obtained for eleven randomly
generated $200 \times 200$ clusters (histogram).
Poisson (dashed line) and Wigner-Dyson (solid line) distributions are shown for
comparison purposes.}
\label{p_rmf_panel}
\end{figure}

\begin{figure}
\begin{picture}(236,230) (0,10)
\epsfbox{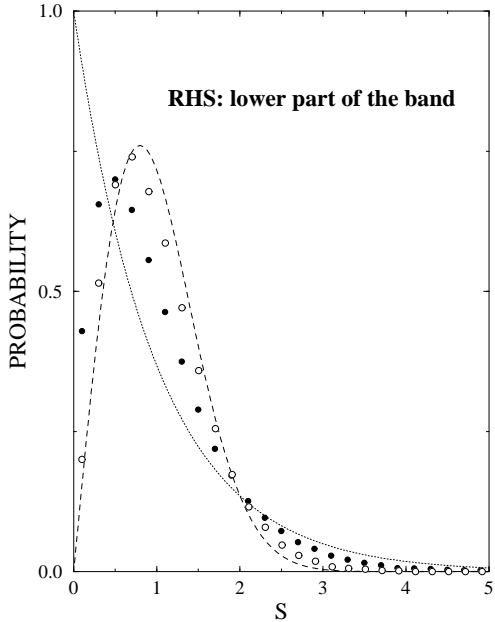}
\end{picture}
\caption{Distribution of nearest neighbor spacings in the lower part of the
band produced by the real version of Hamiltonian (\ref{RMF})
for two different system sizes: open circles
correspond to an ensemble of $12 \times 12$ clusters whereas filled
circles show the distribution for an ensemble of $200 \times 200$ clusters.
Poisson and Wigner GOE distributions are shown for comparison.
\label{p_rhs}}
\end{figure}

Fig.(\ref{p_rmf_panel}) shows the results of a standard analysis
of nearest neighbor spacings produced by the RMF model.
The statistics of small clusters (144 sites) closely follows Wigner surmise
whereas statistics of the largest system that we can analyse (40000 sites)
shows a small but clear deviation from GUE statistics.
On the other hand, larger scaling features can be observed if
the level spacing distribution obtained at the lower part of the
spectrum of the RHS model is analysed (see Fig.(\ref{p_rhs})).
Level statistics for small sizes (empty circles)
reasonably follows Wigner-Dyson distribution (dashed line)
whereas larger sizes give rise to a distribution
(filled circles) that is approaching Poisson distribution (dotted line).

\subsubsection{Scaling of the variance of the spacing distribution}

Since the visual inspection of nearest neighbor spacing distribution is not a
quantitative way of analysing their evolution with system size,
I have selected
the variance of the distribution as the scaling magnitude. Matrices belonging
to the GOE (GUE) give a distribution of variance
equal to $0.286$ ($0.180$)\cite{mehta},
whereas the variance of the Poisson distribution is $1$.
So, a continuous increase of the width of the level spacing
distribution (measured here by its variance) should denote the path
extended-to-localized followed by wavefunctions.
Large samples of nearest neighbor spacings has been obtained by direct
diagonalization of Hamiltonian matrices obtained for random realizations
of disorder. The total number of levels
has been maintained almost constant
($\sim 200 000$ for RHS model and $\sim 480 000$ for RMF model) as the
size of the square cluster has been increased from $12 \times 12$
to $200 \times 200$.

\begin{figure}
\begin{picture}(236,220) (-30,-70)
\epsfbox{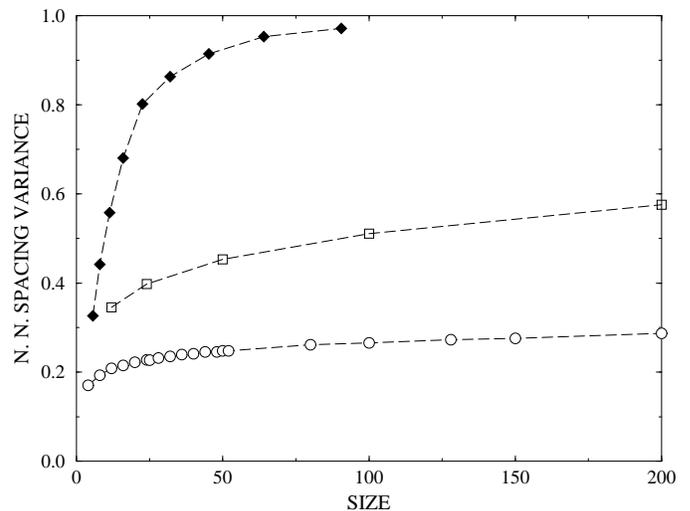}
\end{picture}
\caption{Scaling behavior of nearest neighbor spacing variance as
a function of cluster size. Squares give the evolution of the variance
calculated for the whole band of square clusters of side $L$
for the RHS model, circles represent the same magnitude for the RMF model
and, filled diamonds give the scaling behavior for a $2 \times L$ cluster
of RMF model (in this case the size is not the length of the square side
as before but $\protect \sqrt{2L}$)}
\label{scaling}
\end{figure}

Firstly, I present the results obtained for the variance of the whole
sets of spacings. They are given in Fig.(\ref{scaling})
for both models together
with the results obtained for rectangular clusters (these are the only
results for this cluster shape presented in the paper).
While scaling is obviously towards localization for $2 \times L$ systems,
asymtotic values for squares (RHS model) and circles (RMF model) cannot
be inferred in an unique way. A good fit is obtained using the following
model curve:

\begin{equation}
\begin{array}{c}
\sigma=A+C_1/L+C_2/\log(L)+ \\
      +C_3/(L \log(L))+C_4/L^2+C_5/{\log(L)}^2  ~~~~~~,\\
\end{array}
\label{fit}
\end{equation}

\noindent
where $\sigma$ means variance and $L$ is the cluster size.
Unfortunately, while results for RHS point towards a value $A=1$ in a
three parameter fit (ideal localization of {\it all} eigenstates),
RMF results for $A$ are not so well defined (the number of parameters makes
appreciable differences) but point to an $A$ value smaller than 1 although
a five parameter fit with fixed $A=1$ also goes through all circles.
If this overall fit could be taken quite seriously, it would mean that some
percentage of the total number of states (about fifty percent since $A \approx
0.6$) does not follow a random statistics even in the thermodynamic limit.
I will come back to the discussion of this point after the presentation
of all results.

\begin{figure}
\begin{picture}(236,180) (15,0)
\epsfbox{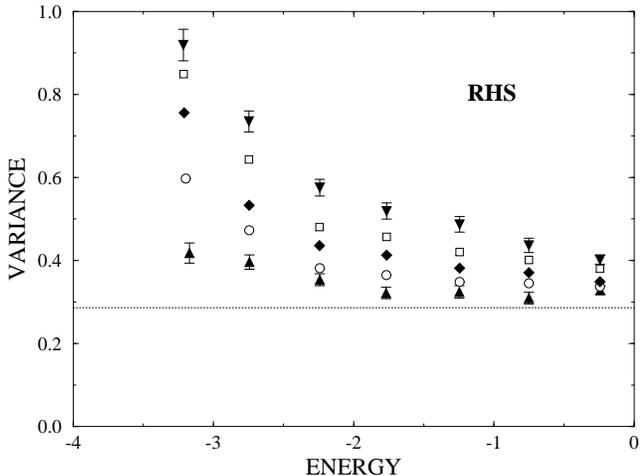}
\end{picture}
\caption{Variance of the nearest neighbor spacing distribution
of the RHS model as a function of the
energy plotted for increasing sizes of the cluster:
the side of the square cluster equals
12, 24, 50, 100, and 200 for up triangles, circles, diamonds, squares,
and down triangles, respectively.
The upper half of the band shows identical behavior.
Error bars are given for the smallest and the largest cluster sizes.
\label{anchura_rhs}}
\end{figure}

Secondly, an analysis that takes into account the energy region from
which level spacings originate has been done. Level spacings have
been grouped in seven sets corresponding to seven energy regions and
the variance of the corresponding distributions has been calculated.
Fig.(\ref{anchura_rhs}) shows the flow towards Poisson statistics
followed by the level statistics within the different energy regions.
Scaling in the lower part of the band shows a rapid tendency to localization
while states near the band center show a much slower ratio.
This behavior easily explains why localization is easily detected
near the band edges but is almost invisible near the band center
(see Fig.(\ref{sites_rhs})).

\begin{figure}
\begin{picture}(236,180) (15,0)
\epsfbox{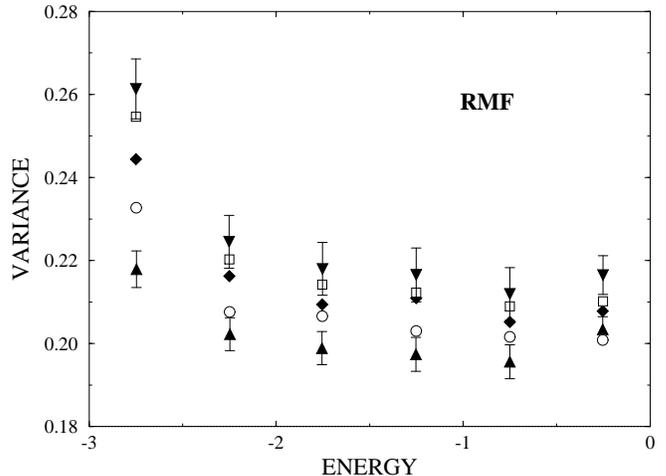}
\end{picture}
\caption{Variance of the nearest neighbor spacing distribution
of the RMF model as a function of the
energy plotted for increasing sizes of the cluster:
the side of the square cluster equals
12, 25, 50, 100, and 200 for up triangles, circles, diamonds, squares,
and down triangles, respectively.
Error bars are given for the smallest and the largest cluster sizes.
\label{anchura_rmf}}
\end{figure}

Numerical analysis is still more difficult for the Hermitian RMF model since
scaling towards localization is much slower for this model.
Remember that the number of level spacings is large and almost
constant ($\sim 480 000$) for all considered system sizes.
In this case, the variance should start close to the
value of $0.180$ that corresponds to the nearest neighbor level spacing
distribution given by matrices of the Gaussian Unitary Ensemble\cite{mehta}.
Fig.(\ref{anchura_rmf}) shows the evolution of the variance of
nearest neighbor level spacings within different energy regions.
As before, states close to the
band edge localize at small sizes ($\leq 200$) but the body of wavefunctions
shows a much slower scaling behavior (compare the scales of y-axis in Figs.
(\ref{anchura_rhs}) and (\ref{anchura_rmf})).
Since the increase of the variance in a given energy range is roughly
the same for the four increases of the cluster side, a logarithmic dependence
of the variance with the system size could be inferred (Notice that the number
of sites is increased by a factor of four in each step). If this were the case,
that would imply that only huge clusters would show exponential localization
of wavefunctions, and consequently, Poisson statistics of level spacings.

In conclusion, although the scaling law at the studied sizes seems to be
logarithmic for the main part of the band,
numerical results show without doubt that the whole spectrum is scaling
towards larger values of the variance, and therefore, towards localization.
In any case, level statistics of both RHS and RMF models {\it is not scaling}
towards the corresponding Wigner-Dyson statistics. The only alternative to
a complete scaling towards a random distribution of levels, i.e., towards
exponential localization of the whole spectrum, is the existence of some
new non-universal statistics for this Hamiltonian models. This possibility
will be further discussed in the last Section.

\subsubsection{Number variance statistics}

Number variance statistics measures fluctuations in the number of levels
that appear in an interval of fixed length. In this case, the statistical
variable is $n$ (number of states in an interval of renormalized length
equal to $s$) and its variance $\langle {(n-<n>)^2} \rangle$ is calculated.
Wigner-Dyson ensembles produce spectra of remarkable rigidity which in
turn means small values of the number variance\cite{mehta}.
On the other hand, a random sequence of levels gives rise to a number variance
equal to the length of the interval.
The advantage of using this statistics to characterize a level sequence is
that long range correlations are explicitly shown.
Number variance statistic is also closely related with the
variance of the distance of one level to its n-th neighbor\cite{brody}.

In accordance with the analysis done in previous subsections, number
variance shows an increasing separation of RHS (RMF) statistics from
Wigner-Dyson GOE (GUE) values as the size of the system increases.
Since it is really not plausible
that this tendency could change beyond some critical size,
we can conclude that the overall spectra flows towards Poisson statistics,
i.e., towards localization for both RHS and RMF models.
Although not shown in the paper, when the band is analysed grouping levels
within different energy regions,
the result of previous subsubsection is recovered:
scaling towards random statistics of states close to the band edges is
quite rapid whereas scaling is much slower near band center.
Once again, the quasi
mobility edge defined by results given in Figs.(\ref{sites_whole_rhs}),
(\ref{sites_whole_rmf}) and (\ref{dimension})
signalizes the change of behavior.

\section{Discussion}

The discussion of the results presented in Section III can be done
according to two different scenarios.
Within the first one, the existence of only
two fixed points for the flow of eigenstates of a matrix belonging to one
of the universality classes is assumed. In our case, since matrices
describing RMF model (Eq.(\ref{RMF})) are Hermitian (the system is not
invariant under time reversal), the two fixed points are:
(i) the one described by the Gaussian Unitary Ensemble (i.e., Wigner-Dyson
statistics of levels sequence, Porter-Thomas statistics for wavefunctions,
etc.) and (ii), the one corresponding to localization (Poisson statistics for
the random sequence of levels, increment of the number of sites having
a vanishing wavefuntion weight, etc.). If this assumption is true, all
numerical results prove overall scaling towards localization.
Moreover, the same trend has been proved for states belonging to different
energy regions. Nevertheless, in spite of the clear and extense numerical
evidence supporting this conclusion, one word of caution is necessary.
Since my analysis is exclusively based on the scaling of averaged magnitudes,
there is always the possibility that a countable
number of states of vanishing weight in averages do not follow overall
scaling and remain extended. This conclusion is in agreement
with the widely accepted point of view that
states are localized in infinite two-dimensional systems no matter
how small the degree of disorder is\cite{localizacion}.

\begin{figure}
\begin{picture}(236,350) (-25,-15)
\epsfbox{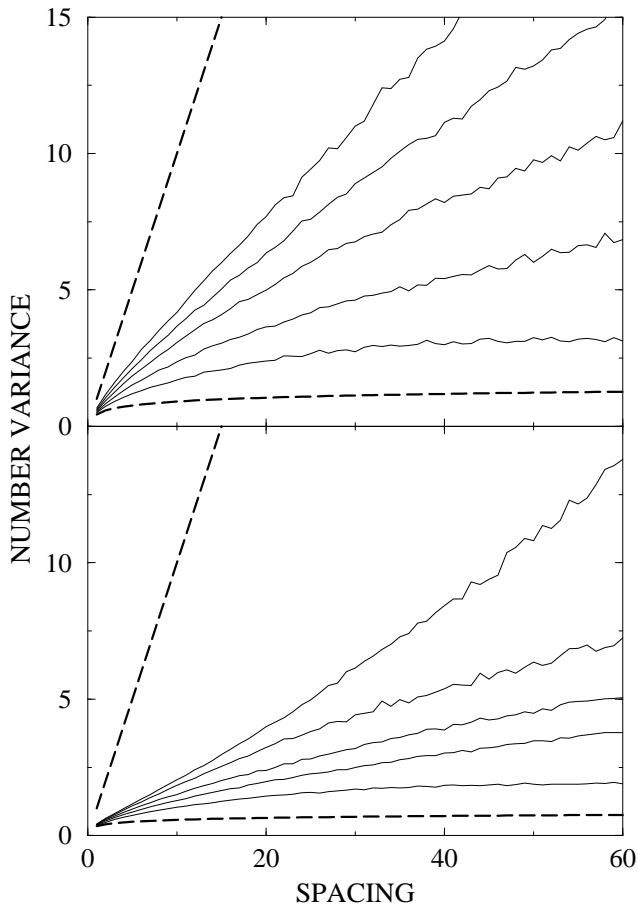}
\end{picture}
\caption{Number variance as a function of spacing
for both RHS model (upper panel) and RMF (lower panel).
Results for $12 \times 12$, $24 \times 24$, $50 \times 50$,
$100 \times 100$ and, $200 \times 200$ square clusters are given as
continuous lines that monotonically go from the corresponding Gaussian
ensemble result (lower dashed line) towards the straigt dashed line
that gives the poissonian statistics corresponding to a random level sequence.
\label{number}}
\end{figure}

On the other hand, a second scenario is also possible. States close
to the band edges are certainly exponentially localized
but it could happen that states
within the band were extended, at least in a loose form:
states whose spatial extension increases as
$ P \sim L^\alpha$, with $\alpha < 2$, cover an infinite number
of sites for $L \rightarrow \infty$ although the percentage of visited
sites relative to the total number of sites vanishes:
$P/L^2 \rightarrow 0$.
In this case, the observed flow towards random (localized) statistics
of all measured statistical magnitudes
could eventually stop at certain "new" fixed point. Under this assumption,
statistics would start at Wigner-Dyson fixed point because small
matrices are {\it almost} filled and separate from this point for
larger matrices. All numerically observed effects would be assign to
finite size effects.

One can give further plausibility to the second option using results obtained
for an ensemble of random {\it real} band matrices.
It has been shown both numerically\cite{casati} and mapping the
problem into a non-linear supersymmetric $\sigma$ model\cite{mirlin_BM} that
the average number of sites visited by a wavefunction is
$\sim b^2$ being $b$ the band width. This gives a typical localization
length of the order of $b$. In our case, we are indeed dealing with band
matrices of a particular kind: Hamiltonian (\ref{RMF}) on a $L \times L$
cluster of the square lattice gives band matrices of width $L$ with only four
nonzero matrix elements within the band. As a consequence, a localization
length $\xi \sim L$ could be expected for RHS model and also,
presumably, for RMF model. If this were the case, I think that such a large
localization length (of the order of the sample size)
can only be understood as signalising some kind of "geometrical" or fractal
localization as the one described in the previous paragraph.

Incidentally, this second scenario would be more interesting from
the point of view of experimental mesoscopic physics than the canonical one.
Instead of measuring
{\it universal} statistics like Wigner surmise corresponding to the GUE,
a {\it new} although non-universal statistics could be obtained (Compare, for
example, Wigner surmise in the bottom panel of Fig.(\ref{p_rmf_panel}) with
the distribution corresponding to $40 000$ sites described by the RMF model).

In summary, there are unfortunately two different but internally consistent
explanations of the whole set of numerical results. Either the number of fixed
points is limited to the well known universality classes of non-interacting
systems and then random magnetic fluxes imply localization at all
energies or a new fixed point exists for the particular type of random
band matrices ensemble considered in this paper and then the major 
part of eigenstates would exhibit fractal behavior instead of
standard exponential localization.

\acknowledgments
I gratefully acknowledge the hospitality of the Condensed Matter Theory
Group of M.I.T. where this work was begun and 
people of the Applied Physics Department of the University of Alicante
where the work was almost finished.
I thank E. Mucciolo for giving me the program that obtains the level spacing
distribution from a set of eigenenergies and M. Ortu\~no for insisting about
the necessity of using alternative ways to characterize localization.
This work has been partially supported by
spanish CICyT (grant MAT94-0058-C02).


\begin{references}
\bibitem{fractal}
I speak about fractal states in a loose form, using this adjective for
states in energy regions at which the mean participation ratio
follows an $L^\alpha$ law
with $\alpha$ smaller than the space dimension (2 in this context).
Although strict fractal states certainly follow this law, self-similarity
is not necessarily implied by a power law increase of the average number
of sites covered by states about some fixed energy.
\bibitem{ls}
Level  statistics of a system belonging to the universal orthogonal class
would move towards the statistics corresponding to the Gaussian Orthogonal
Ensemble (GOE) whereas level statistics of a system of the universal unitary
class would move towards the statistics described by the Gaussian
Unitary Ensemble (GUE).
\end{references}
\end{document}